# Pseudogap Formation in the Nodal-Line Semimetal NaAlGe

Takahiro Yamada[1], Daigorou Hirai[2], Tamio Oguchi[3], Hisanori Yamane[1], and Zenji Hiroi[2]

[1]*Institute of Multidisciplinary Research for Advanced Materials, Tohoku University, Katahira 2–1–1, Aoba-ku, Sendai 980–8577, Japan*

[2]*Institute for Solid State Physics, University of Tokyo, Kashiwanoha 5–1–5, Kashiwa, Chiba 277–8581, Japan*

[3]*Center for Spintronics Research Network, Osaka University, Toyonaka, Osaka 560-8531, Japan*

NaAlSi and NaAlGe are isostructural and isoelectronic semimetals with topological nodal lines close to the Fermi level. Despite having virtually identical electronic structures, NaAlSi exhibits superconductivity below $T_c$ = 6.8 K, whereas NaAlGe does not. We investigate NaAlGe by measuring its electrical resistivity, Hall effect, magnetic susceptibility, and heat capacity using single crystals. It is revealed that NaAlGe is not a simple semimetal but rather has an unusual ground state with a small pseudogap of ~100 K close to the Fermi level. We argue that the formation of the pseudogap in NaAlGe is due to an unexpected Fermi surface instability, such as an excitonic instability, as opposed to the electron–phonon instability that leads to the formation of the superconducting gap in NaAlSi.

## 1. Introduction

Topological materials with band structures that contain linearly crossing band dispersion have attracted attention. Nodal-line semimetals with extended band crossing points along a specific line have been emphasized theoretically[1,2] and realized in actual materials such as ZrSiS,[3,4] PbTaSe$_2$,[5-7] and CaAgP.[8,9] Here, we will focus on the nodal-line semimetals NaAlSi and NaAlGe. Both crystallize in anti-PbFCl type layered structures (space group $P4/nmm$),[10] in which the Al and Si/Ge atoms are covalently bonded to form conducting layers that are separated by layers of ionic Na atoms. According to first-principles electronic state calculations, NaAlSi and NaAlGe are isoelectronic quasi-two-dimensional semimetals with highly dispersive electron-like Al 3$s$ bands and less dispersive hole-like Si 3$p$ (Ge 4$p$) bands.[11,12] The two bands cross to generate complex nodal-line structures near the Fermi level in the absence of spin–orbit interactions (SOI).[12-15]

NaAlSi exhibits superconductivity below $T_c$ = 7 K.[16] The superconductivity is most likely of the conventional $s$-wave type based on phonon-mediated Cooper pairing;[14,17-20] However, recent study utilizing single crystals demonstrated that fractional superconductivity could exist at magnetic fields greater than the bulk upper critical field, which was suggested to originate from surface states.[21] Despite their similar electronic structures, NaAlGe was reported to be a non-superconductor above 2 K,[16] which was recently confirmed by resistivity and magnetic susceptibility measurements.[22] This led to the hypothesis that superconductivity is mediated by low-mass, high-frequency phonons of Si rather than Ge.[18] Nevertheless, a recent study on complete solid solutions NaAlSi$_{1-x}$Ge$_x$ using a series of single crystals demonstrated that as the Ge content increases, the $T_c$ does not decrease smoothly but drops abruptly at $x \sim 0.45$, which may be difficult to explain in terms of a gradual decrease in phonon energy with increasing $x$.[22] This fact suggests that something unforeseen occurred on the Ge side. NaAlGe's fundamental properties have been largely unexplored due to its chemical instability and lack of single crystals.

In this study, we report unexpected electronic properties for NaAlGe. While first-principles calculations predict a semimetallic band structure very similar to that of NaAlSi, bulk measurements using single crystals reveal the formation of a pseudogap of approximately 100 K magnitude. It is hypothesized that NaAlGe conceals a specific Fermi surface instability, such as an excitonic instability, whereas NaAlSi conceals an electron–phonon instability.

## 2. Experimental

Single crystals of NaAlGe were grown using the Na–Ga flux method, as previously described.[20,22] A mixture of Na : Al : Ge : Ga = 3 : 1 : 1 : 0.5 was placed in a boron nitride crucible under an argon atmosphere and sealed in a stainless-steel reaction container. The container was heated to 1073 K in an electric furnace and then slowly cooled to 823 K for 80–100 hours. Single crystals with a platelet shape and a 1–4 mm edge were extracted from the crucible after heating the crucible to 573 K under vacuum and holding it for 10 hours to evaporate the excess Na. Because the resulting single crystals were extremely hygroscopic (much more so than NaAlSi), all measurements were conducted as quickly as possible in a dry atmosphere.

To refine the crystal structure and composition, a single-crystal X-ray diffraction (XRD) measurement was performed on the crystal using Mo-K$\alpha$ radiation ($\lambda$ = 0.71075 Å) in a diffractometer (Bruker AXS, D8 QUEST). The APEX3 software package was used both for data collection and unit cell refinement. The observed data were corrected for multiscan absorption using the SADABS program,[23] and the crystal structure was refined using the SHELXL-2018/3 program installed in the WinGX software.[24,25] In addition, the crystal's chemical composition was determined using wavelength-dispersive X-ray (WDX) spectroscopy in an electron probe microanalyzer system (JEOL XA-8200).

Measurements of resistivity ($\rho$), Hall coefficient ($R_H$), and heat capacity ($C$) were carried out in a physical property measurement system (PPMS, Quantum Design Inc.). Resistivity was measured using the standard four probe method with indium metal electrodes on the crystal's top surface; silver or gold metal pastes could not be used as electrodes due to their high contact resistance. Magnetic susceptibility ($\chi$) was measured in a 7 T magnetic field applied along the $a$ axis in a magnetic property measurement system (MPMS, Quantum Design Inc.).

First-principles electronic structure calculations of NaAlSi and NaAlGe were performed by using the all-electron full-potential linearized augmented plane wave (FLAPW) method[26-28] implemented in the HiLAPW code[29] on the basis of the generalized gradient approximation to the density-functional theory (DFT).[30] The SOI was self-consistently taken into account for the valence and core states by the second variation scheme.[31] The energy cutoffs for wavefunction and potential expansions were 20 and 160 Ry, respectively. Γ-centered 16 × 16 × 8 mesh $k$ points were used to sample the Brillouin zone with the tetrahedron integration scheme in the self-consistent calculations.[32] The density of states (DOS) was calculated using finer 64 × 64 × 36 mesh $k$ points. The irreducible representations were extracted from the eigenfunctions by considering the double-group of $k$ to plot the band structure with the proper crossing and anti-crossing.

## 3. Results
### 3.1 Crystal structure and chemical composition

The single crystal structural refinement yielded in the following crystallographic parameters (Table 1): $a$ = 4.1634(2) Å, $c$ = 7.4146(4) Å, $z$(Na) = 0.63552(11), and $z$(Ge) = 0.21252(2),[22] which are compared to the previously reported values; $a$ = 4.164(8) Å, $c$ = 7.427(9) Å, $z$(Na) = 0.618(5), and $z$(Ge) = 0.217(3).[10] The band structure was calculated using these parameters.

The average chemical composition of five crystals, as determined by the WDX spectroscopy, was $Na_{1.13}Al_{0.97}Ga_{0.01}Ge_{0.89}$, indicating that approximately 10% excess Na was substituted for Ge and ~1% Ga remained at the Al site from the flux; for NaAlSi and Si-rich solid solutions, obtained crystals were nearly stoichiometric with 1–2% Ga contamination.[20,22] However, an XRD structural refinement based on the assumption of a mixture of Na and Ge atoms at the Ge site revealed that there was no such substitution of Na for Ge: the occupancy of the Ge atom was 1.002(3). This indicates that a nearly stoichiometric sample was obtained. The cause for the compositional deviation in the chemical analysis is unknown, but it may be related to the crystal's hygroscopic nature: in fact, Na-rich phases formed quickly on the crystal's surface as a result of the reaction with water vapor, and the WDX spectroscopy was somehow surface sensitive.

### 3.2 Calculated band structures

Our fully-relativistic DFT FLAPW calculations of the electronic structure for NaAlSi and NaAlGe yielded nearly identical band dispersions and DOS profiles (Fig. 1), which are almost consistent with previous findings.[11-15] The Al 3$s$ electron band and Si 3$p$ or Ge 4$p$ hole bands are located close to the Fermi energy ($E_F$) in both band structures. They display high levels of dispersion around the Γ point in the plane and become relatively flat along the Γ–Z axis, indicating a quasi-two-dimensional feature. Near the Fermi level, the two bands intersect to form complex nodal lines.[12-15] However, SOI causes anti-band crossing, resulting in the formation of small gaps with energies less than 5 meV in NaAlGe[12] and greater than 20 meV in NaAlSi.[3,33] The SOI also causes a greater splitting of the hole bands for Ge 4$p$ than Si 3$p$.

The DOS close to $E_F$ in Fig. 1(c) is dominated by heavy Si/Ge hole bands; however, the transport properties may be governed by light electrons. DOS values at $E_F$ are 1.428 and 1.169 states per eV per unit cell for NaAlSi and NaAlGe, respectively. Apparently, the reduced DOS of NaAlGe is due to the increased SOI splitting of the Ge bands. Our DOS values are considerably greater than the previous values of ~1.0 states per eV per unit cell.[11,17] The difference may be due to differences in the structural parameters used in the calculations as well as the precision of the calculations; we used 10659 $k$ points within the irreducible Brillouin zone in our calculations, which is significantly more than 2772 $k$ points reported in the literature.[17] On the basis of the band structure calculations, one would anticipate that the two compounds have comparable electronic properties.

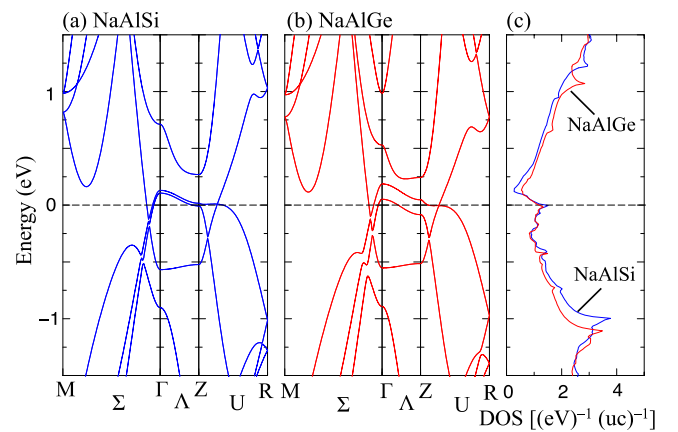

**Fig. 1.** (Color online) Calculated band dispersions for (a) NaAlSi and (b) NaAlGe, as well as (c) their density of states (DOS) close to the Fermi level ($E$ = 0). For the calculations, the structural parameters obtained from the present single crystal XRD analyses (Table 1) were utilized. Their DOS profiles are virtually identical, especially near the Fermi level.

### 3.3 Transport properties

The in-plane resistivities $\rho$ of NaAlSi and NaAlGe are compared in Fig. 2(a). Resistivity values are comparable at 300 K, 1.7 and 1.5 mΩ cm, respectively. When NaAlSi is cooled, $\rho$ decreases to ~0.1 mΩ cm and becomes zero below $T_c$ = 6.8 K due to the superconducting transition; the relatively large $\rho$ value above $T_c$ may be due to a strong scattering by electron–phonon interactions as well as impurity scattering by Ga atoms. In stark contrast, the $\rho$ of NaAlGe increases below 200 K to 23.5 mΩ cm at 2 K, following metallic behavior near room temperature; there is no indications of a superconducting transition above 2 K. Thus, the ground state of NaAlGe must differ significantly from that of metallic and superconducting NaAlSi.

NaAlGe exhibits a relatively modest increase in resistivity in comparison to conventional insulators with well-defined band gaps. The Arrhenius plot of Fig. 2(b) demonstrates a nonlinear but upward concave dependence. Attempting to fit the curve to a line at the low-temperature limit is only possible for data below 3 K, resulting in an absurdly small gap of 0.2 K. Thus, there is no charge gap. Alternatively, one can fit the data below 5 K to a logarithmic form, $-\ln T$, as would be expected for Kondo scattering by a magnetic impurity,[34] but such magnetic impurities are not included in NaAlGe. Therefore, this particular scattering is irrelevant.

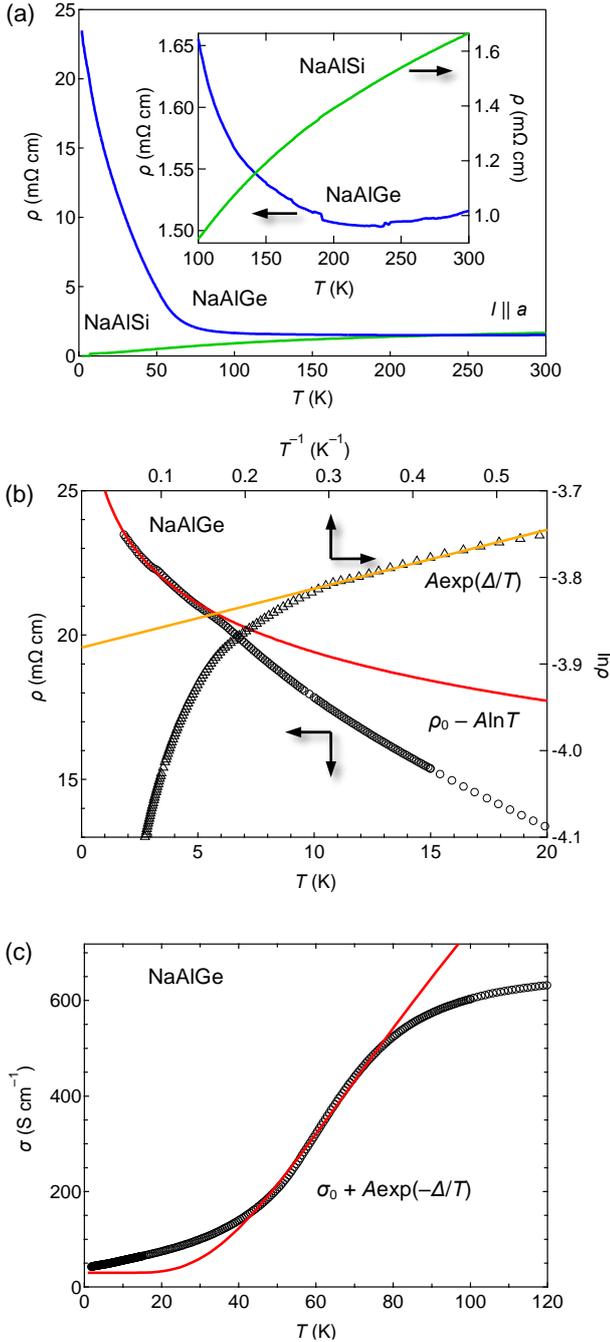

Fig. 2. (Color online) (a) Resistivities $\rho$ of NaAlSi and NaAlGe measured with an electric current along the $a$ axis in the plane. (b) $\rho$ versus $T$ plot (left–bottom axes) and $\ln\rho$ versus $T^{-1}$ plot (right–top axes) for NaAlGe. The line on the $\rho$–$T$ plot follows the form $\rho_0 - A\ln T$ ($\rho_0 = 0.02501(4)$ mΩ cm, $A = 0.00243(3)$ mΩ cm), while the line on the $\rho$–$T^{-1}$ plot follows the form $A\exp(\Delta/T)$ ($A = 0.02072(4)$ mΩ cm, $\Delta = 0.232(4)$ K). (c) Conductivity of NaAlGe below 120 K. The line is an approximation of the form $\sigma_0 + A\exp(-\Delta/T)$ ($\sigma_0 = 30$ S cm$^{-1}$, $A = 2800(50)$ S cm$^{-1}$, $\Delta = 136(1)$ K).

The temperature dependences of carrier density $n$ as determined by Hall effect measurements are shown in Fig. 3. As with NaAlSi,[20] the Hall resistivity $\rho_{xy}$ is proportional to the applied magnetic field along the $c$ axis over the entire temperature range. This suggests that light electrons dominate the transport, while heavy holes play a secondary role. The variation of the slope indicates that $n$ is $1$–$2 \times 10^{21}$ cm$^{-3}$ at high temperatures for both compounds, which are in good agreement with the calculated values of $1$–$1.3 \times 10^{21}$ cm$^{-3}$ for NaAlSi.[11] However, $n$ decreases slightly below 50 K for NaAlSi and more steeply below 100 K for NaAlGe at low temperatures. The latter's $n$ becomes almost constant at $2.3 \times 10^{20}$ cm$^{-3}$ below 50 K, which is approximately one order of magnitude less than at 300 K. Thus, a significant change in the electronic structure must occur at around $T^* \sim 100$ K. This not-so-small $n$ value and its saturating temperature dependence imply that a metallic state is maintained even after the transformation.

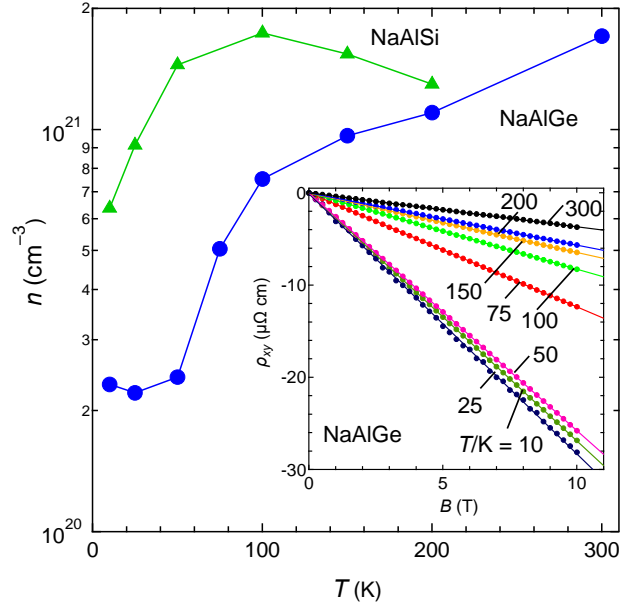

Fig. 3. (Color online) Temperature dependences of carrier density $n$ determined by Hall effect measurements for NaAlSi and NaAlGe. The inset depicts the Hall resistivity of NaAlGe at various temperatures as a function of magnetic field. The carrier density is determined using linear fits and the presumption that electron carriers are predominant.

### 3.4 Thermodynamic properties

The temperature dependences of magnetic susceptibility $\chi$ are compared for the two compounds in Fig. 4. NaAlSi exhibited a high degree of anisotropy in its $\chi$:[20] $\chi_c$ ($B \parallel c$) was significantly less than $\chi_a$ ($B \parallel a$) due to the large contribution of Landau orbital diamagnetism, which was estimated to be $-2.81 \times 10^{-5}$ cm$^3$ mol$^{-1}$ at 300 K as the difference between $\chi_c$ and $\chi_a$. Thus, in Fig. 4, the $\chi_a$ data are compared after the nuclear diamagnetic contribution has been subtracted, which can be scaled to the Pauli paramagnetic response $\chi_P$.

Both magnetic susceptibilities increase similarly from similar values at 300 K upon cooling. This temperature dependence is ascribed to a band smearing effect: when a sharp peak in DOS exists just below $E_F$, as shown in Fig. 1, thermally excited quasiparticles attain a smaller DOS at higher energy than at $E_F$, thereby reducing the average DOS and thus $\chi_P$ at elevated temperatures. The saturation observed for NaAlSi near $T = 0$ is naturally explained by this line, and the value at the lowest temperature, $6.7 \times 10^{-5}$ cm$^3$ mol$^{-1}$, should correspond to the Pauli paramagnetic contribution of the ground state if the small Curie tail coming from defects or others is ignored. Compared to the calculated $\chi_P$ value of $2.4 \times 10^{-5}$ cm$^3$ mol$^{-1}$ derived from the DOS, the experimental $\chi_P$ value is 2.8 times greater.

In contrast, the $\chi$ of NaAlGe exhibits a distinct decrease below 80 K, which appears to begin at 100 K. This strongly suggests that, below $T^*$, not only does band smearing occur, but also a substantial reduction in DOS. A similar decrease in magnetic susceptibility, known as the pseudogap phenomenon, has been observed in the high-$T_c$ cupric oxide superconductors;[35] for example, in $YBa_2Cu_3O_{6.7}$, the $^{63}Cu$ spin–lattice relaxation rate and Knight shift decrease below 100 K.[36] The reduction in NaAlGe is relatively modest. If the intrinsic $\chi_P$ is set to the minimum value of $6.1 \times 10^{-5}$ cm$^3$ mol$^{-1}$ at 30 K, there is a significant enhancement of 3.2 over the calculated value of $1.9 \times 10^{-5}$ cm$^3$ mol$^{-1}$ for NaAlGe, which is comparable to the enhancement for NaAlSi.

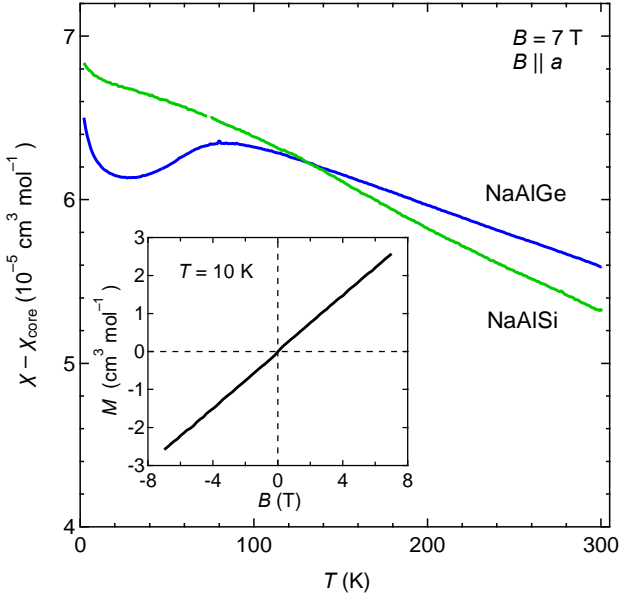

**Fig. 4.** (Color online) Magnetic susceptibility for NaAlSi and NaAlGe after subtracting the nuclear diamagnetic contribution $\chi_{core}$; $\chi_{core} = -0.85 \times 10^{-5}$ and $-1.4 \times 10^{-5}$ cm$^3$ mol$^{-1}$, respectively. A 7 T magnetic field was applied along the $a$ axis, which may have eliminated the Landau orbital diamagnetic contribution. The inset depicts a magnetization versus magnetic field plot at 10 K, which is nearly linear in $B$ but has a small nonlinear component due to free spins, as indicated in the main panel by the Curie tail appearing below 30 K.

Figure 5 depicts the low-temperatures heat capacity $C$. There is no anomaly for NaAlGe that indicates a phase transition. The fact that the $C$ of NaAlGe increases more rapidly than that of NaAlSi indicates a softer lattice: the slopes in the $C/T$ versus $T^2$ plot indicate that the Debye temperature of NaAlGe is 235 K, which is less than that of NaAlSi, which is 370 K.[20] In addition, the intersections of the plots reveal that the Sommerfeld coefficients $\gamma$, which are proportional to the DOS at $E_F$, are 0.45 and 2.15 mJ K$^{-2}$ mol$^{-1}$ for the Ge and Si compounds, respectively. Thus, although the DOS of Ge is reduced by a factor of 5, it remains greater than zero within the experimental resolution, indicating the presence of a metallic ground state. The band structure calculations yield $\gamma$ values of 1.4 and 1.7 mJ K$^{-2}$ mol$^{-1}$ for NaAlGe and NaAlSi, respectively. Thus, NaAlSi has a mass enhancement factor of 1.3, while NaAlGe has a mass reduction of 32%. All the determined quantities are summarized in Table 1.

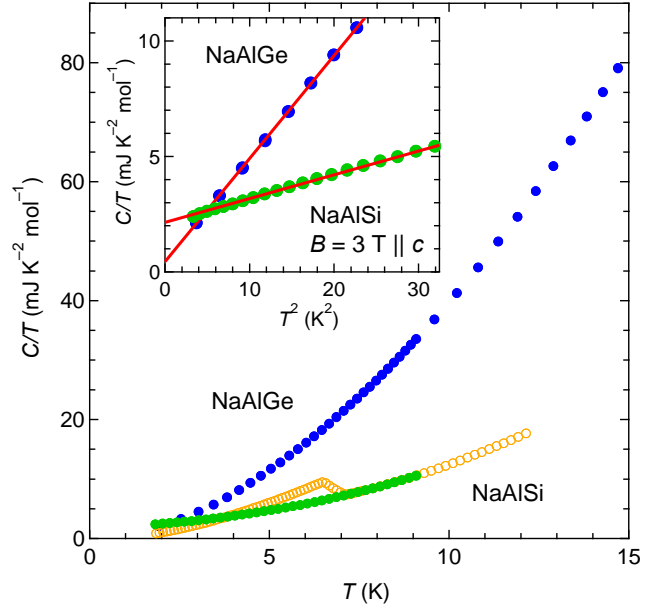

**Fig. 5.** (Color online) Heat capacity divided by temperature at zero magnetic field for NaAlGe and NaAlSi. For NaAlSi, another data taken at a magnetic field of 3 T along the $c$ axis are also shown, where the superconducting transition is completely suppressed.[20] The inset depicts a plot of $C/T$ versus $T^2$ for Ge data at zero field and Si data at $B = 3$ T. The lines are fits to the form $C/T = \gamma T + \beta T^3$, which results in 0.45(2) mJ K$^{-2}$ mol$^{-1}$ and 0.447(1) mJ K$^{-4}$ mol$^{-1}$ for NaAlGe, and 2.15(1) mJ K$^{-2}$ mol$^{-1}$ and 0.1031(4) mJ K$^{-4}$ mol$^{-1}$ for NaAlSi.

**Table 1.** Crystallographic and physical parameters of NaAlSi[20] and NaAlGe.[22]

|  | NaAlSi | NaAlGe |
|---|---|---|
| $a$ [Å] | 4.12170(10) | 4.1634(2) |
| $c$ [Å] | 7.3629(2) | 7.4146(4) |
| $z$(Na) | 0.63461(8) | 0.63552(11) |
| $z$(Si/Ge) | 0.20764(4) | 0.21252(2) |
| DOS [states (eV)$^{-1}$ (uc)$^{-1}$] | 1.428 | 1.169 |
| HT resistivity [mΩ cm] | 1.7 (300 K) | 1.5 (300 K) |
| LT resistivity [mΩ cm] | 0.17 (10 K) | 23.5 (2 K) |
| Carrier density [cm$^{-3}$] | $6.2 \times 10^{20}$ (10 K) | $2.3 \times 10^{20}$ (10 K) |
| $\chi_P^{exp}$ [cm$^3$ mol$^{-1}$] | $6.7 \times 10^{-5}$ | $6.1 \times 10^{-5}$ |
| $\chi_P^{calc}$ [cm$^3$ mol$^{-1}$] | $2.4 \times 10^{-5}$ | $1.9 \times 10^{-5}$ |
| $\gamma_{exp}$ [mJ K$^{-2}$ mol$^{-1}$] | 2.15 | 0.45 |
| $\gamma_{calc}$ [mJ K$^{-2}$ mol$^{-1}$] | 1.7 | 1.4 |
| $R_W$ | 2.3 | 10 |
| Ground state | Superconductivity | Pseudogap |
| Charac. temp. [K] | 6.8 | ~100 |
| Fermi surface instability | Electron–phonon int. | Electron–hole (exciton) int.? |

## 4. Discussion
### 4.1 Pseudogap formation in NaAlGe

Our transport and thermodynamic measurements indicate that NaAlGe is neither a simple semimetal nor an insulator, but is approaching a gapped state at low temperatures. Below 100 K, resistivity begins to rise, and carrier density begins to fall. However, unlike an insulator, resistivity does not diverge toward $T = 0$ and has a much weaker temperature dependence, and the carrier density remains finite at $T = 0$. Moreover, magnetic susceptibility decreases dramatically below 100 K, and the Sommerfeld coefficient is significantly smaller than that of NaAlSi but still finite. All of these observations are consistent with the existence of a 100 K pseudogap in NaAlGe, which was not predicted by band structure calculations; the introduction of SOI has little effect on the creation of the pseudogap.[3,12,33]

At elevated temperatures above 100 K, on the other hand, both compounds display nearly identical behavior: similar magnitudes and temperature dependences in resistivity, carrier density, and magnetic susceptibility. As a result, both compounds share the nearly identical high-temperature states. At low temperatures, a 100 K pseudogap opens only in NaAlGe when a specific fluctuation develops that is not taken into account in the band structure calculations.

Due to the pseudogap opening, the electrical conductivity shown in Fig. 2(c) decreases with decreasing thermally excited carriers as the temperature decreases but remains a finite value $\sigma_0$ at $T = 0$ due to the finite DOS. The $\sigma_0$ value must be less than 43 S cm$^{-1}$ at 1.8 K. In Fig. 2(c), the conductivity data are fitted to the form $\sigma_0 + A\exp(-\Delta/T)$. A fit to the data between 40 and 80 K with $\sigma_0$ set to 30 S cm$^{-1}$ yields $A = 2800(48)$ S cm$^{-1}$ and $\Delta = 136(1)$ K; the value of $\Delta$ is insensitive to the choice of $\sigma_0$. Thus, the decrease in conductivity is consistent with a pseudogap opening of magnitude of ~100 K. With this $\sigma_0$ value, a small electron mobility of 1.3 cm$^2$ (Vs)$^{-1}$ is estimated based on the carrier density of $2.3 \times 10^{20}$ cm$^{-3}$ at 10 K.

Let us compare the values of $\gamma$ and $\chi_P$ for the two compounds. The $\gamma$ value of NaAlGe is reduced by a factor of five compared to that of NaAlSi, while the $\chi_P$ value is reduced by only 10 percent. Due to the fact that both quantities are proportional to the DOS at $E_F$ in the free-electron model, this indicates a selective enhancement of $\chi_P$, as in the case of strongly correlated electron systems.[37] The Wilson ratio $R_W$, denoted by $(\pi^2/3)(k_B/\mu_0)^2(\chi/\gamma)$, is 2.3 for the Si compound and 10 for the Ge compound. The former value is typical for strongly correlated electron systems and has been suggested to relate to the characteristic saddle-shaped hole bands in NaAlSi.[20] By contrast, the latter is massive. Even when ambiguity in estimating the $\chi_P$ value is considered, the $R_W$ value cannot be reduced significantly. Therefore, NaAlGe has an atypical ground state characterized by a large decrease in $\gamma$ and an increase in $\chi_P$, in comparison to NaAlSi and the calculated band structure. While the reason for this unexpected increase in magnetic susceptibility is unknown, it must convey critical information regarding the ground state of NaAlGe.

One may question how the non-stoichiometric chemical composition affects the electronic properties; the WDX analysis showed a composition of approximately Na$_{1.1}$AlGe$_{0.9}$, whereas the crystal structure analysis revealed a virtually stoichiometric composition. When the rigid band picture is maintained, however, the off-stoichiometry-induced shift in $E_F$ in the DOS profile shown in Fig. 1 obviously does not result in a pseudogap opening. On the other hand, the observed resemblance in resistivity, carrier density, and magnetic susceptibility between the two compounds at temperatures above $T^*$ indicates that they have a minor difference in their electronic structure. As a result, we can reasonably conclude that our observations are inherent to NaAlGe.

*4.2 Excitonic instability?*

Let us briefly discuss the origin of the pseudogap in NaAlGe. The origin is evidently related to a certain Fermi surface instability that is not included in the band structure calculation. In high-$T_c$ cupric oxide superconductors, complex instabilities involving spin, charge, and lattice degrees of freedom may play a crucial role in the formation of the most well-known pseudogap.[35] In the $sp$ electron system of NaAlGe, strong electron correlations and magnetic instabilities are ruled out, as is charge-density wave instability, due to the absence of nesting in the Fermi surfaces and structural instability. Alternatively, we consider the possibility of excitonic instability in the nodal-line semimetal.

In semiconductors with very narrow band gap or semimetals with a small band overlap and a low carrier density, it is well established that an unscreened Coulomb interaction will cause an electron and a hole to exist in a bound state.[38,39] When such excitons undergo Bose–Einstein condensation, an exotic phase known as the "excitonic insulator" is formed. Several compounds, such as 1T-TiSe$_2$,[40] Ta$_2$NiSe$_5$,[41,42] and ZrSiS,[3,43-45] have been studied in this context. However, the excitonic insulator is an elusive phase of matter, as it is always challenging to distinguish between the exotic insulator and a trivial band insulator.[42]

ZrSiS has a similar two-dimensional crystal structure (PbFCl type) and nodal-line semimetallic band structure to the present compounds.[3,43] Although theoretical predictions of excitonic instability and a transition to an excitonic phase have been made,[44,45] experimental evidence has remained inconclusive. Interestingly, at sufficiently low temperatures, an associated pseudogap formation has been predicted,[45] but this excitonic pseudogap has not yet been detected using angle-resolved photoemission spectroscopy.[33] ZrSiS could be in the vicinity of a quantum phase transition to an excitonic insulator.[45]

It would be intriguing to determine if our observations of the pseudogap in NaAlGe are related to the predicted excitonic instability for ZrSiS. Notably, the majority of candidates are electronic insulators, placing them in the semiconductor regime for excitonic instability.[39] Comparatively, semimetallic ZrSiS and NaAlGe are located in the semimetal regime, which requires a low carrier density and consequently weak screening to enhance the exciton pairing interaction. ZrSiS, which has small Fermi surfaces and a low carrier density, can achieve these conditions, whereas NaAlGe, which has large Fermi surfaces and a high carrier density, cannot. Consequently, NaAlGe may be a poor candidate, if viewed naively. However, judgement should be withheld until additional theoretical considerations on the electronic state of NaAlGe and the possibility of Fermi surface instability have been completed.

One simple question is why, despite their comparable electronic structures, only NaAlGe exhibits excitonic instability and not NaAlSi. Ge's more polarizable 4$p$ electrons are thought to provide a higher level of screening than Si's 3$p$ electrons; therefore, NaAlGe is less favorable for exciton formation than NaAlSi. Perhaps this effect is negligible in context of current systems. The underlying excitonic instability has most likely been eliminated in NaAlSi due to the strong electron–phonon coupling that causes the superconductivity, whereas the inherent excitonic instability develops in NaAlGe due to the weaker electron–phonon coupling. In other words, NaAlGe is not a superconductor because excitonic instability toward the formation of electron–hole pairs outweighs phononic instability toward the formation of electron–electron pairs. Notably, even in NaAlSi, a pseudogap opening tendency is detected prior to the superconducting transition: as shown in Fig. 3, the carrier density decreases slightly as the temperature falls below 50 K. It would be interesting to explore the properties of NaAlSi when the electron–phonon interaction is suppressed, as when pressure is applied.

We believe that the physics of NaAlSi and NaAlGe is intriguing. To address the preceding question and to comprehend the pseudogap's origin and significance for excitonic instability in NaAlGe, additional theoretical considerations and experimental efforts are required.

## 5. Conclusions

The transport and thermodynamic properties of the nodal-line semimetal NaAlGe were investigated and compared to those of the isoelectronic NaAlSi. NaAlGe forms a pseudogap below 100 K, whereas NaAlSi forms a superconducting gap below 6.8 K, despite the fact that their calculated band structures and physical properties at elevated temperatures are equivalent. In comparison to the electron–phonon instability in NaAlSi, we suggest that excitonic instability-related fluctuations are responsible for the pseudogap formation in NaAlGe. The pseudogapped ground state in NaAlGe appears to be remarkable, with a substantial increase in Pauli paramagnetic susceptibility and an enormous Wilson ratio.

**Acknowledgments**

The authors are grateful to S. Uji for valuable comments. This research was financially supported by JSPS KAKENHI Grants (JP20H02820 and JP20H05150).